\documentclass{article} 
\usepackage{conference,times}

\usepackage{amsmath,amsfonts,bm}

\def\eqref#1{equation~\ref{#1}}

\def\1{\bm{1}}

\DeclareMathAlphabet{\mathsfit}{\encodingdefault}{\sfdefault}{m}{sl}
\SetMathAlphabet{\mathsfit}{bold}{\encodingdefault}{\sfdefault}{bx}{n}

\usepackage{hyperref}
\usepackage{url}
\usepackage{booktabs}
\usepackage{tikz}
\usepackage{cleveref}
\usepackage{xcolor}
\usepackage{mdframed}
\usepackage{subcaption}
\usepackage{filecontents}
\usepackage{wrapfig}
\usepackage{soul}
\usepackage{pgfplots}

\usepackage{algpseudocode}
\usepackage{algorithm}
\usepackage{optidef}
\usepackage{cleveref}
\usepackage{amsmath}
\usepackage{amssymb}
\usepackage{graphicx}

\usepackage{tikz}

\pgfplotsset{compat=1.15}
\usepgfplotslibrary{fillbetween}

\definecolor{cs1}{HTML}{c7e9b4}
\definecolor{cs2}{HTML}{7fcdbb}
\definecolor{cs3}{HTML}{41b6c4}
\definecolor{cs4}{HTML}{1d91c0}
\definecolor{cs5}{HTML}{225ea8}
\definecolor{cs6}{HTML}{253494}
\definecolor{cs7}{HTML}{081d58}

\definecolor{cd2}{HTML}{addd8e}
\definecolor{cd3}{HTML}{78c679}
\definecolor{cd4}{HTML}{41ab5d}
\definecolor{cd5}{HTML}{238443}
\definecolor{cd6}{HTML}{006837}
\definecolor{cd7}{HTML}{004529}
\definecolor{ForestGray}{HTML}{004529}

\definecolor{codegreen}{rgb}{0,0.6,0}
\definecolor{codegray}{rgb}{0.5,0.5,0.5}
\definecolor{codepurple}{rgb}{0.58,0,0.82}
\definecolor{backcolour}{rgb}{0.95,0.95,0.92}
\definecolor{jcred}{HTML}{e31a1c}
\definecolor{jcgreen}{HTML}{33a02c}
\definecolor{jcblue}{HTML}{1f78b4}
\definecolor{jcorange}{HTML}{ff7f00}
\definecolor{jcpurple}{HTML}{6a3d9a}
\definecolor{jclightred}{HTML}{fb8072}
\definecolor{jclightgreen}{HTML}{b3de69}
\definecolor{jclightblue}{HTML}{80b1d3}
\definecolor{jclightorange}{HTML}{fdb462}
\definecolor{jclightpurple}{HTML}{bebada}
\definecolor{jcredl}{HTML}{fb8072}
\definecolor{jcgreenl}{HTML}{b3de69}
\definecolor{jcbluel}{HTML}{80b1d3}
\definecolor{jcorangel}{HTML}{fdb462}
\definecolor{jcpurplel}{HTML}{bebada}
\definecolor{jcyellow}{HTML}{fdb462}

\usepackage{pifont}  

\definecolor{jcgreen}{HTML}{33a02c}
\definecolor{jcred}{HTML}{e31a1c}

\newcommand{\cmark}{\textcolor{jcgreen}{\ding{51}}}%
\newcommand{\xmark}{\textcolor{jcred}{\ding{55}}}%

\colorlet{definitioncolour}{ForestGray!10!white}

\mdfdefinestyle{fancyenv}{
    skipabove=0.8\baselineskip,
    skipbelow=0.\baselineskip,
    innertopmargin=8pt,
    innerbottommargin=8pt,
    linewidth=2pt,
    topline=true,
    frametitleaboveskip=\dimexpr-\ht\strutbox\relax,
    nobreak=true,
}

\newcounter{definition}

\title{AMPLE: Event-Driven Accelerator for Mixed-Precision Inference of Graph Neural Networks}

\author{
Pedro Gimenes~\thanks{Corresponding author.}, Yiren Zhao, George Constantinides \\
Department of Electrical \& Electronic Engineering,
Imperial College London \\
\texttt{\{pedro.gimenes19, a.zhao, g.constantinides\}@ic.ac.uk} \\
}

\iclrfinalcopy
\begin{document}

\maketitle

\begin{abstract}
Graph Neural Networks (GNNs) have recently gained attention due to their performance on non-Euclidean data. The use of custom hardware architectures proves particularly beneficial for GNNs due to their irregular memory access patterns, resulting from the sparse structure of graphs. However, existing FPGA accelerators are limited by their double buffering mechanism, which doesn't account for the irregular node distribution in typical graph datasets. To address this, we introduce \textbf{AMPLE} (Accelerated Message Passing Logic Engine), an FPGA accelerator leveraging a new event-driven programming flow. We develop a mixed-arithmetic architecture, enabling GNN inference to be quantized at a node-level granularity. Finally, prefetcher for data and instructions is implemented to optimize off-chip memory access and maximize node parallelism. Evaluation on citation and social media graph datasets ranging from $2$K to $700$K nodes showed a mean speedup of $243\times$ and $7.2\times$ against CPU and GPU counterparts, respectively.
\end{abstract}

\section{Introduction} \label{sec:intro}

Graphs serve as powerful representations for capturing relationships between entities, which are represented as nodes, connected together by edges. This structure enables modeling a wide range of complex systems, including social networks \citep{Alamsyah2021SocialRepresentation}, biological interactions \citep{Wu2021GraphChallenges}, and recommendation systems \citep{Wang2021GraphReview}. Graph Neural Networks (GNNs) have emerged as a transformative approach for processing graph data, designed to learn from complex relational information by exploiting the interconnections within the graph \citep{Kipf2016Semi-SupervisedNetworks, Velickovic2017GraphNetworks}. 

Inference on GNN models can be divided into two main computational phases, (1) \textbf{Aggregation} and (2) \textbf{Transformation} \citep{gilmer2017neural}. In the Aggregation phase, a permutation-invariant function such as summation or mean is applied over the feature embeddings of a node's neighbors. The results of this phase are then utilized in the Transformation phase, which consists of a fully-connected layer used to generate the updated feature embedding for each node. While the Transformation phase presents a \textit{highly regular computational pattern}, which can be effectively accelerated on a parallelized device such as a GPU, the Aggregation phase involves many irregular memory accesses due to the random and sparse nature of typical graph data. Additionally, aggregation latency is a function of a node's degree, which follows a highly non-uniform distribution. As such, an efficiently-designed GNN accelerator needs to alleviate the computational irregularity of the Aggregation phase while leveraging the regularity of the Transformation phase \citep{Yan2020HyGCN:Architecture}.

\begin{table*}[t]

\caption{Summary of graph processing features across hardware platforms. Although CPU parallelization is possible, multi-threaded CPUs have limited core count compared to parallelized accelerators. Event-driven programming is possible in GPUs, however computation does not follow a node-centric flow. Although pre-fetching is possible in GenGNN, all incoming messages for nodes in flight are required to be stored on-chip. Finally, no existing accelerators support arbitrary multi-precision computation.}
\centering

\resizebox{\textwidth}{!}{ 
\begin{tabular}{ccccc}
\toprule
\textbf{Hardware Platform} &
  \textbf{Parallelization} &
  \textbf{Event-Driven Programming} &
  \textbf{Node Pre-Fetching} &
  \textbf{Multi-Precision} \\
\midrule
CPU (Intel Xeon)   & \xmark & \cmark & \xmark & \xmark \\
GPU (RTX A6000)    & \cmark & \cmark & \xmark & \xmark \\ 
HyGCN \citep{Yan2020HyGCN:Architecture} & \cmark & \xmark & \xmark & \xmark \\ 
GenGNN \citep{Abi-Karam2022GenGNN:Acceleration} & \cmark & \xmark & \cmark & \xmark \\ 
\textbf{\textbf{AMPLE}} & \cmark & \cmark & \cmark & \cmark \\
\bottomrule
\vspace{-0.5cm}
\end{tabular}%
}
\label{tab:competitors}
\end{table*}



Although CPU memory systems are a mature and highly optimized technology, the sparse structure of graph data renders traditional cache systems less effective, since node aggregation incurs a high number of accesses to non-contiguous memory ranges. Inference on GPUs offers higher performance due to the deep level of parallelism, however, these devices are limited by high-latency memory management mechanisms. Additionally, there is no support for inter-phase pipelining, meaning aggregation results must be stored into off-chip memory before being re-fetched for the transformation phase. Finally, modern devices have limited support for computation with low-precision numerical formats.

These considerations have motivated the design of several GNN accelerators. HyGCN leverages a set of Single Instruction Multiple Data (SIMD) cores for aggregation, and a systolic array for node transformation \citep{Yan2020HyGCN:Architecture}. Meanwhile, GenGNN was proposed as a model-agnostic framework for GNN acceleration, addressing the gap between the development pace of GNN models and custom accelerators \citep{Abi-Karam2022GenGNN:Acceleration} through High-Level Synthesis tools. Table \ref{tab:competitors} summarizes the characteristics of available GNN hardware platforms. Despite the benefits of previously proposed GNN accelerators, (i) the double-buffering mechanism deployed in HyGCN is not well suited for graph computation due to the non-uniform distribution of node degrees. Under this paradigm, low degree nodes must wait for higher degree nodes before computation can proceed, causing a high number of pipeline gaps. This highlights the need for an \textbf{event-driven programming flow}, where nodes are independently allocated resources and scheduled onto the accelerator. 
Additionally, (ii) neither accelerator offers hardware support for model quantization. As observed by Tailor \textit{\emph{et al.}} \citep{Tailor2020Degree-Quant:Networks}, the accuracy cost of quantization in GNNs is predominantly due to the aggregation phase and directly correlated to a node's degree. As such, casting low-degree nodes to lower-precision formats while preserving high-degree nodes in high precision leads to reduced memory cost and resource usage at a low cost to model accuracy. 
Finally, (iii) existing accelerators require on-chip buffering of node embeddings for the entire input graph. As such, these have limited applicability for inference on large graphs ($>100k$ nodes) where embeddings cannot feasibly be stored on-chip, highlighting the need for a \textbf{node-centric pre-fetching system} to hide memory access latency while the accelerator is busy.





We address these shortcomings by introducing a novel GNN accelerator, AMPLE, contributing the following:

\begin{itemize}
     \item We showcase an event-driven programming model for GNN acceleration, by enabling the host to program nodes asynchronously through memory-mapped registers. 
    
    \item We propose an architecture featuring a heterogeneous pool of multi-precision Aggregation Cores connected through a Network-on-Chip, which are dynamically allocated to nodes according degree and precision.
    
    
    \item We evaluate AMPLE on large-scale graph datasets ranging from 2K to 700K nodes, achieving an average speedup of $243\times$ and $7.2\times$ compared to CPU and GPU baselines, respectively. 
\end{itemize}

The body of this paper is structured as follows. Section~\ref{sec:background} covers background on GNNs and neural network quantization. Section~\ref{sec:arch} explains the architecture of the AMPLE accelerator, including how each high-level feature is achieved at the circuit level. Finally, Section~\ref{sec:results} explains the testing methodology and experimental results against CPU/GPU baselines.

\section{Background}
\label{sec:background}

\subsection{Graph Representation}

A graph $G = (\mathcal{V}, \mathcal{E})$ is a set of nodes/vertices $\mathcal{V}$ and edges $\mathcal{E}$. The set of feature representations at layer $l$ is denoted by matrix $X^{(l)} \in \mathcal{R}^{N \times D}$, where $N = |\mathcal{V}|$ is the number of nodes and $D$ is the feature size. An element $e_{i, j} = (v_i, v_j)$ present in $\mathcal{E}$ indicates that there is a connection between nodes $v_i$ and $v_j$, meaning node $v_j$ is contained in the set of $i$'s neighbors, $\mathcal{N}_i$, and $v_i$ is contained in $\mathcal{N}_j$. In an undirected graph, the edge element $e_{i, j}$ corresponds to $e_{j, i}$. The connections in a graph can be represented using an $N \times N$ adjacency matrix, where each element $A_{i, j}$ represents an edge between nodes $i$ and $j$.




\subsection{Graph Neural Networks (GNNs)}

Within a GNN, graph data is transformed over several layers to perform classification and/or regression tasks on the full graph or individual nodes/edges. GNNs can be represented through the Message Passing Mechanism \citep{gilmer2017neural}, which generalizes the node update law as follows.
\begin{equation} \label{eq:message_passing}
    \mathbf{x}_i^{l+1} = \gamma (\mathbf{x}_i^l, \mathcal{A}_{j \in \mathcal{N}(i)}(\phi(\mathbf{x}_i^l, \mathbf{x}_j, e_{i, j}^l)))
\end{equation}

It can be seen that in the general case, each node aggregates incoming messages represented as an arbitrary function $\phi$, which is equivalent to aggregating neighboring embeddings when $\phi = \mathbf{x}_j^l$. Messages are aggregated through an arbitrary permutation-invariant aggregation function $\mathcal{A}_{j \in \mathcal{N}(i)}$ over the neighborhood of node $i$, and and an arbitrary transformation function $\gamma (\mathbf{x}_i^l, \mathbf{m}_i^l)$, where $\mathbf{m}_i^l$ is the result of aggregation (i.e.~$\mathbf{m}_i = \mathcal{A}_{j \in \mathcal{N}(i)} \phi(\mathbf{x}_i^l, \mathbf{x}_j^l, e_{i, j}^l)$).

\subsubsection{Graph Convolutional Networks (GCN)}

GCNs emerged as a solution analogous to Convolutional Neural Networks in the computer vision domain \citep{Kipf2016Semi-SupervisedNetworks}. The element-wise node update law for a single GCN layer is shown in Equation \ref{eq:gcn}.
\begin{equation} \label{eq:gcn}
    \mathbf{x}_i^{l+1} = W \left ( \sum_{j \in \mathcal{N}_i \cup \{ i \}} \frac{e_{j,i} }{\sqrt{\hat{d}_j \hat{d}_i}} \mathbf{x}_j^{l} \right )
\end{equation}

The normalization factors are given by ${\hat{d}_i = 1 + \sum_{j \in \mathcal{N}(i)} e_{j,i}}$. It can be seen that $\mathcal{A}$ is taken as the summation ${\mathcal{A}=\sum_{j \in \mathcal{N}_i} \phi(\mathbf{x}_j, e_{i, j})}$, with ${\gamma (\mathbf{x}_i, \mathbf{m}_i) = W \mathbf{m}_i}$.




\subsubsection{Graph Isomorphism Networks (GIN)}

GIN was proposed as a model that can provably generate distinct feature updates for two graphs that can be shown to be non-isomorphic through the Weisfeiler-Lehman test \citep{Leman2018THERO}, thus maximizing its representational capacity \citep{Xu2018HowPA}. The update law is given by the following, where $\epsilon$ is a small scalar for numerical stability.
\begin{equation} \label{eq:gin}
    \mathbf{x}^{l+1}_i = MLP \left( (1 + \epsilon) \cdot
        \mathbf{x}^{l}_i + \sum_{j \in \mathcal{N}(i)} \mathbf{x}^{l}_j \right)
\end{equation}

The same aggregation $\mathcal{A}$ is used as in GCN. In contrast to GCN, GIN does not make use of normalization factors in aggregation (i.e. ${\phi = \mathbf{x}_j}$), and a residual connection is added after aggregation, which is equivalent to a self-connection in the graph's adjacency matrix.

\subsubsection{GraphSAGE} 

GraphSAGE was proposed as an inductive framework to generate feature embeddings with high representational capacity for unseen nodes and/or sub-graphs \citep{Hamilton2017InductiveRL}.
\begin{equation} \label{eq:sage}
    \mathbf{x}^{l+1}_i = W_1 \mathbf{x}_i + W_2 \cdot \left ( \underset{j \in \mathcal{N}(i)}{\text{mean}} \sigma(W_3 \mathbf{x}_j^{l} + \mathbf{b}) \right )
\end{equation}

It can be seen that the message passing function $\phi$ is taken as a fully-connected layer with activation $\sigma$ over the neighbouring embeddings $\mathbf{x}_j$, $\mathcal{A}$ is taken as the mean, and the transformation ${\gamma (\mathbf{x}_i, \mathbf{m}_i) = W_1 \mathbf{x}_i + W_2 \mathbf{m}_i}$ where $W_1, W_2$ are linear projection matrices. The projection parameterized by $W_1$ can be seen as a scaled residual connection.

\subsection{Neural Network Quantization}

Quantization has been widely explored as a method for reducing model complexity and computational latency in neural networks. Quantization-Aware Training (QAT) enables minimizing accuracy loss at low-precision representations by quantizing activations in the forward pass, making use of the Straight-Through Estimator (STE) in the backwards pass to estimate the non-differentiable quantization gradients. In general, activations are quantized following Equation \ref{eq:quantization}, where $q_{min}, q_{max}$ form the chosen range of representable values, $s$ is the scaling factor to place $x$ into the required range, $z$ is the zero-point (floating point equivalent of the value 0 in the quantized space) and the brackets represent the rounding operation.
\begin{equation}\label{eq:quantization}
    x_q = \min (q_{max}, \max(q_{min}, \left\lfloor \frac{x}{s} + z \right\rceil))
\end{equation}

The $\min$ and $\max$ functions are in place to show that any values beyond the specified range assume the fixed-point value at the limit. Following this, activation can be de-quantized by ${\hat{x} = (x_q - z)s}$, where $\hat{x}$ is an approximation of the original floating-point value. 

\subsubsection{Quantization-Aware Training for Graph Neural Networks} \label{section:degreequant}

Degree-Quant, proposed by Tailor \emph{et al.}, was one of the first approaches applying Quantization-Aware Training to Graph Neural Networks \citep{Tailor2020Degree-Quant:Networks}. Firstly, Tailor \emph{et al.} suggest that the aggregation phase of GNN inference is the predominant source of quantization error. This effect is observed more heavily in nodes with higher in-degrees, which can be intuitively understood since the absolute magnitude of aggregation grows with the number of neighbors. The growth in expected aggregation for high-degree nodes affects the $q_{max}$ and $q_{min}$ values, reducing the quantization resolution due to these outliers in the distribution of aggregation results.

The authors of Degree-Quant address the issue of quantization error by stochastically applying a protection mask at each layer following the Bernoulli distribution \citep{Tailor2020Degree-Quant:Networks}. Protected nodes operate in floating-point, while non-protected nodes operate in fixed-point. A node's probability of protection is a function of its degree, interpolated within a parametrizable range $ {[} p_{min}, p_{max} {]}$, where the graph nodes with minimum/maximum neighbor counts are assigned the limit probabilities. 


\section{Architecture} \label{sec:arch}
\begin{figure}
    \centering
    \includegraphics[width=0.55\linewidth]{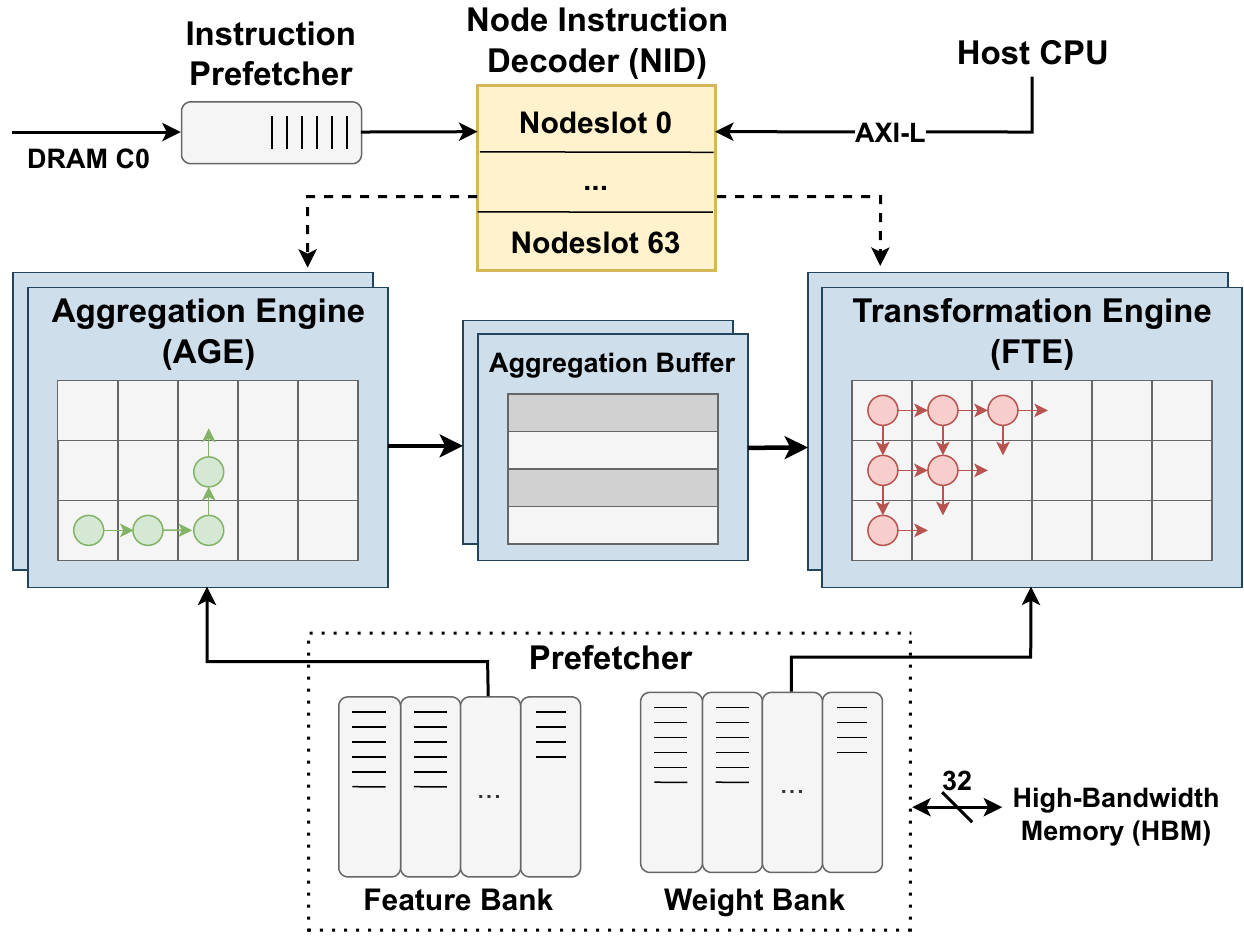}
    \caption{\textbf{AMPLE} Top Level Diagram. Packets propagate through dimension-order routing in the Aggregation Engine's Network-on-Chip (shown in green), and are driven diagonally into the Transformation Engine's systolic array (shown in red). Dashed lines represent control flow interfaces, while solid lines represent data flow between units. Node embeddings are fetched through HBM, while instructions are stored in DRAM.}
    \label{fig:fc_base_top}
\end{figure}
As shown in Figure \ref{fig:fc_base_top}, AMPLE is composed of the following functional units, with their respective functions.

\begin{itemize}
\item \textbf{Node Instruction Decoder (NID)}: communication with the host device and driving other functional units to schedule work onto the accelerator.

\item \textbf{Prefetcher}: fetching and storing layer weights and neighbouring feature embeddings into local memories (the Weight Bank and Feature Bank, respectively).

\item \textbf{Aggregation Engine (AGE)}: performing permutation-invariant aggregation functions over all neighbours of a node through a Network-on-Chip architecture.

\item \textbf{Aggregation Buffer (ABF)}: storage element containing aggregated feature embeddings generated by the AGE.

\item \textbf{Feature Transformation Engine (FTE)}: computing the updated feature embeddings for each node by performing a matrix multiplication between weights in the Weight Bank and aggregation results in the Aggregation Buffer.

\end{itemize}
\subsection{Event-Driven Programming through the Node Instruction Decoder} \label{section:node_scoreboard} 

\begin{table*}[t]
\caption{Example state of the Node Scoreboard at runtime, with nodeslots are found in various states. The adjacency list and updated feature pointers indicate the memory address from which to fetch the list of neighbouring node IDs and write updated feature embeddings, respectively. The precision field dictates which arithmetic cores are allocated at runtime.}
\centering
\resizebox{\textwidth}{!}{ 
\begin{tabular}{ccccccc}
\toprule
\textbf{Slot} &
  \textbf{Node ID} &
  \textbf{Precision} &
  \textbf{State} &
  \textbf{Neighbors} &
  \textbf{Adjacency List Pointer} &
  \textbf{Updated Feature Pointer} \\
\midrule
0 & 267 & float & Transformation & 32 & 0x3BC90188 & 0x4FE8B774 \\
1 & 268 & float & Aggregation & 8 & 0xCAF5C03F & 0xE672109F \\ 
\dots & \dots & \dots & \dots & \dots & \dots & \dots \\
63 & 330 & int4 & Prefetch & 1 & 0x78E26A27 & 0xA4D89ED9 \\ 
\bottomrule
\vspace{-0.5cm}
\end{tabular}%
}
\label{tab:controller_sb}
\end{table*}

Communication between AMPLE and the host is handled by the Node Instruction Decoder (NID), which is a memory-mapped register bank comprised of a configurable number of nodeslots. As shown in Table \ref{tab:controller_sb}, each nodeslot contains the information required to perform a node's aggregation and transformation steps, and a state machine is maintained indicating each node's state. Algorithm \ref{alg:layer_config} shows how work can be offloaded by the host, which runs concurrently with the accelerator. First, the NID is programmed with a number of global and layer-wise parameters, including node/feature counts and aggregation functions. Subsequently, the host programs the nodeslots and updates values in the mask ${available\_nodeslots \in \{0, 1\}^n}$ where $n$ is the number of nodeslots. While a node is programmed, the accelerator performs aggregation and transformation over previously-programmed nodes. The $available\_nodeslots$ mask is then deasserted independently by the accelerator when the computation is finished. Thus, the accelerator supports a node-wise, event-driven computation paradigm. Note that $'1$ and $'0$ indicate a mask full of ones and zeros, respectively.

\begin{algorithm}
\caption{Host programming pseudocode}
\label{alg:layer_config}
\begin{algorithmic}
\Require global parameters $\mathcal{P}$, layers $\mathcal{L}$, nodes $\mathcal{V}$

\State nid\_register\_bank.global\_parameters $\gets$ $\mathcal{P}$
\State available\_nodeslots $\gets$ '1

\For{layer in $\mathcal{L}$}
    \State nid\_register\_bank.layer\_config $\gets$ layer
    \State layer.prefetch\_layer\_weights()

    \While{$\mathcal{V} \neq \emptyset$}
        \If {available\_nodeslots != '0}
            \State chosen$\_$nodeslot $\gets$ choose{(}available\_nodeslots{)}
            \State chosen\_nodeslot.programming $\gets$ $\mathcal{V}$.pop\_head()
            \State available\_nodeslots {[}chosen\_nodeslot{]} $\gets$ 0
        \EndIf
        \EndWhile

        \EndFor

    \end{algorithmic}
\end{algorithm}
After a nodeslot is programmed, the NID then drives the Prefetcher, AGE and FTE to perform the computation, and updates the node's internal state machine after each functional step. No further intervention is required from the host, and an interrupt is sent after step 7 to indicate the nodeslot can be reused. It should be noted that the order in which nodes are programmed within the nodeslots does not imply any priority or time correlation. Typical graph datasets often display high variance in execution time per node, depending on neighbour count and numerical precision. Whenever a nodeslot finishes its computation, it can be immediately reprogrammed by the host with the next node. This event-driven control flow requires the host to run concurrently with the accelerator to monitor its state and drive further work when resources are available. Within the NID, nodes running concurrently are serviced with round-robin arbitration to grant access to shared resources within the Aggregation and Transformation Engines.










\subsection{Mixed-precision Arithmetic} \label{section:age}

\begin{figure*}[t]
    \centering
    \includegraphics[width=0.9\linewidth]{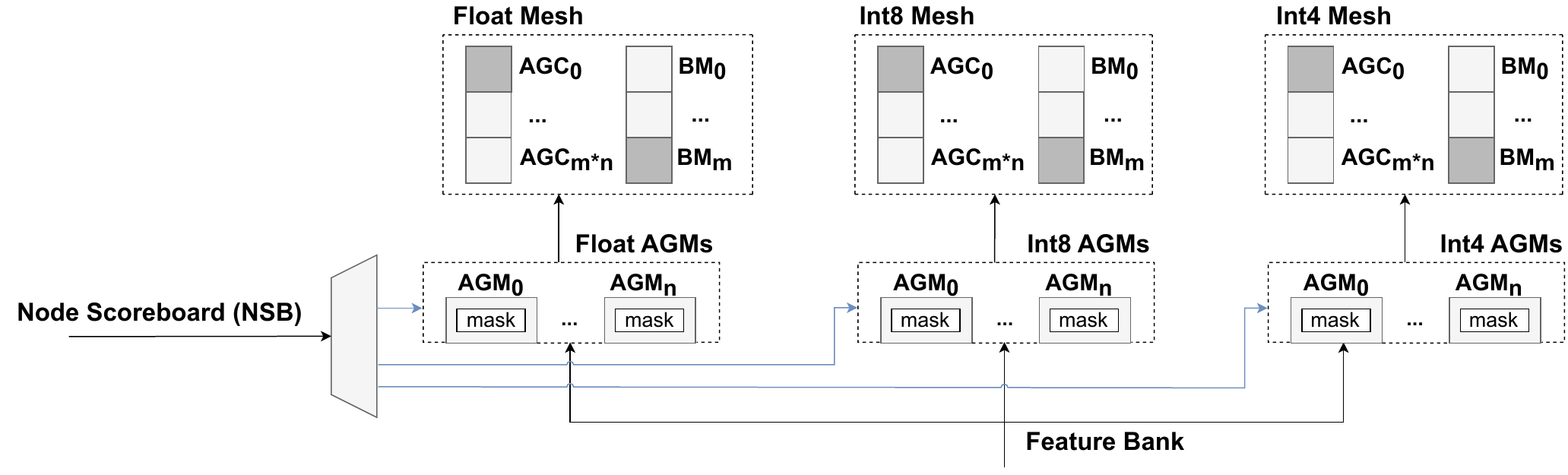}
    \caption{Microarchitecture of AGE configured with three supported precisions. NID requests drive the Aggregation Managers (AGMs), which receive fetched embeddings from the Feature Bank (See Figure \ref{fig:fc_base_top}). These are then transferred to the Aggregation Cores (AGCs) through the network. Aggregation results are then buffered by the Buffering Managers (BMs).}
    \label{fig:aggregation_engine_uarch}
\end{figure*}

Computing units within the AGE and FTE are locally homogeneous, meaning each processing element supports a single numerical precision. Within the Aggregation Engine, these are arranged in a Network-on-Chip (NoC) architecture comprising a heterogeneous grid of processing elements, where the ratio of PEs allocated to each precision can be configured at compile time according to the application requirements. Each PE is coupled to a router responsible for transferring packets over the network. Each packet is comprised of a head flit carrying routing payloads, an arbitrary number of body flits carrying data, and a tail flit. Since there is no requirement for communication between PEs of different precisions, these are placed within isolated sub-networks as shown in Figure \ref{fig:aggregation_engine_uarch}, which acts to reduce packet congestion.

As discussed in Section \ref{sec:intro}, static pipelining through the double buffering mechanism leads to pipeline gaps when computing over graphs with high variance in node degree, since low-degree nodes must wait for high-degree nodes to release resources. This is alleviated in the AGE by dynamically allocating processing elements within each aggregation sub-network according to a node's feature count and precision. As such, nodeslots are allocated resources independently of any other ongoing workload, and these resources can be immediately reused upon completion, thus forming an event-driven programming model. 

\subsection{Large Graph Processing} \label{section:prefetcher_uarch}

\begin{figure}[t!]
    \centering
    \includegraphics[width = 0.5\linewidth]{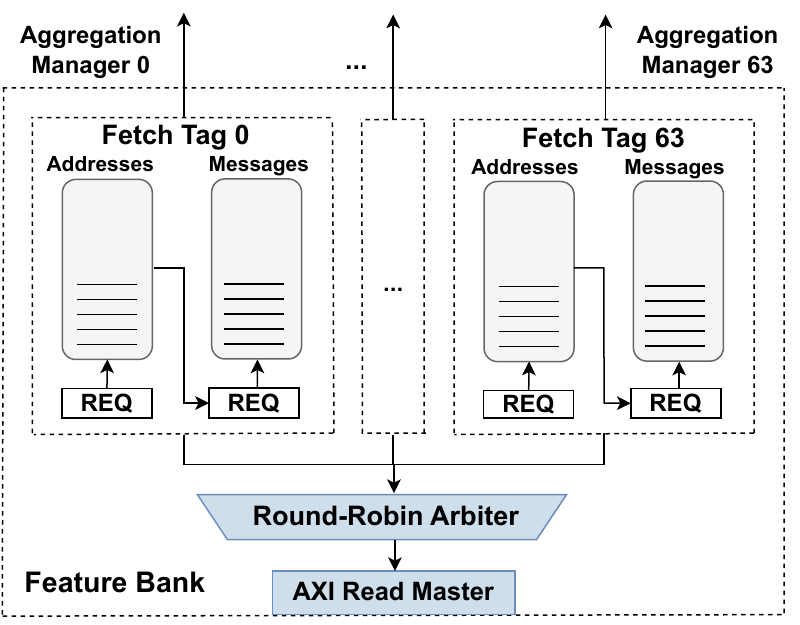}
    \caption{Fetch Tags in the Feature Bank make concurrent memory access requests in a two-stage process; First, the list of neighbouring node IDs is stored in the Address Queue, and these are then used as pointers for the neighbouring feature embeddings, which are stored in the Message Queue.}
    \label{fig:uarch_prefetch}
\end{figure}

Inference over large graphs is enabled by the Feature Bank in the Prefetcher, which contains a storage element named ``Fetch Tag" for each nodeslot in the NID. A group containing a parametrizable number of Fetch Tags is coupled to each HBM bank on the Alveo U280 card, meaning up to 32 Fetch Tags can access memory resources concurrently, alleviating the inherent memory boundedness associated with sparse graph data. Within each Fetch Tag group, access to the HBM bank is granted using round robin arbitration.


The Feature Bank supports the large graph use case via its \textbf{partial response mechanism}. For nodes with degree higher than the Fetch Tag capacity, the Fetch Tag fills the Message Queue and directly unblocks the AGE to begin the aggregation process. Once aggregation begins, the Fetch Tag is re-enabled and continues to fetch the remaining neighbours, hiding the memory access latency. This mechanism leads to lower storage requirement per nodeslot, allowing a higher number of Fetch Tags in the Feature Bank, i.e. deeper node parallelism.


\vspace{-0.1cm}
\section{Experimental Results} \label{sec:results}

Three foundational GNN models were deployed for evaluating the accelerator, with varied architectures, as shown in Table \ref{tab:models}. See Section \ref{sec:background} for each model's update laws. Furthermore, 6 graph datasets were chosen, the first three being small citation networks, and the last three being larger social media graphs. Table \ref{tab:datasets} shows the node count and mean node degree for each evaluated dataset - the latter acts as an indicator of graph sparsity, with an inverse relationship between sparsity and mean degree.

\begin{table}[t]
\caption{GNN models used for benchmarking the accelerator. A residual connection denotes the addition of the node's original embedding after the aggregation or transformation steps.}
\label{tab:models}
\centering
\begin{tabular}{ccccc}
\toprule
 \textbf{Model} &
 \textbf{Aggregation} &
 \textbf{Residual} &
 \textbf{Normalization} \\

\midrule

\textbf{GCN} &
    sum &
    \xmark &
    aggregation \\

\textbf{GIN} &
    sum &
    aggregation &
    \xmark \\

\textbf{GraphSAGE} &
    mean &
    transformation &
    transformation \\

\bottomrule
\end{tabular}
\end{table}

\begin{table}[t]
\caption{Datasets used for benchmarking. DQ ratio shows the ratio of nodes mapped to float precision by the DegreeQuant algorithm, with the rest running in int8.}
\label{tab:datasets}
\centering
\begin{tabular}{cccccc}
\toprule

&
 \textbf{Name} &
 \textbf{Nodes} &
 \textbf{Mean Degree} &
 \textbf{Features} &
 \textbf{DQ Ratio} \\

\midrule

\textbf{CR} &
    Cora &
    2,708 &
    3.9 &
    1,433 &
    2.1 \% \\

\textbf{CS} &
    CiteSeer &
    3,327 &
    2.7 &
    3,703 &
    2.7 \% \\

\textbf{PB} &
    PubMed &
    19,717 &
    4.5 &
    500 &
    2.9 \% \\

\textbf{FL} &
    Flickr &
    89,250 &
    10.0 &
    500 &
    0.2 \% \\

\textbf{RD} &
    Reddit &
    232,965 &
    99.6 &
    602 &
    2.7 \% \\

\textbf{YL} &
    Yelp &
    716,847 &
    19.5 &
    300 &
    0.4 \% \\


\bottomrule
\end{tabular}
\end{table}

\begin{figure*}[t]
    \centering
    \includegraphics[width=\linewidth]{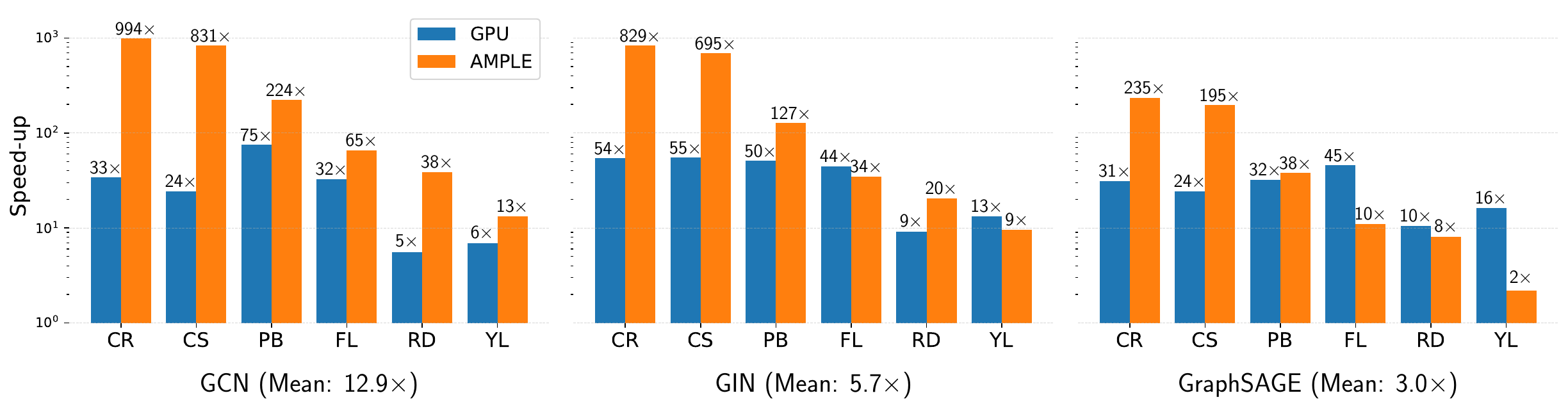}
    \caption{Inference speedup compared to Intel Xeon CPU baseline obtained on the RTX A6000 GPU and AMPLE simulation. The GPU shows an average speedup of $29.8\times$, $37.8\times$ and $26.7\times$ across all datasets for GCN, GIN and GraphSAGE, respectively. Equivalent speedups on AMPLE were $361.3\times$, $285.8\times$ and $81.7\times$. \vspace{-0.5cm}}
    \label{fig:speedup}
\end{figure*}



\begin{table*}[b]
\caption{Inference time for evaluated datasets using a single-layer GCN model. Mean latency is reported over 100 iterations.}
\centering
\resizebox{\textwidth}{!}{ 
    \begin{tabular}{ccccccccccc}
    \toprule
        &
        \multicolumn{2}{c}{\textbf{CPU (Intel Xeon)}} &
        &
        \multicolumn{2}{c}{\textbf{GPU (RTX A6000)}} &
        &
        \multicolumn{4}{c}{\textbf{AMPLE @200MHz}} \\
    
    \cmidrule{2-3}
    \cmidrule{5-6}
    \cmidrule{8-11}
        
    &
        \textbf{Mean} &
        \textbf{Throughput} &
        &
        \textbf{Mean} &
        \textbf{Throughput} &
        &
        \textbf{Mean} &
        \textbf{Throughput} &
        \textbf{Latency} &
        \textbf{Latency} \\
    
    &
        \textbf{Latency {[}ms{]}} &
        \textbf{$[$nodes/ms$]$} &
        &
        \textbf{Latency {[}ms{]}} &
        \textbf{$[$nodes/ms{]}} &
        &
        \textbf{Latency {[}ms{]}} &
        \textbf{{[}nodes/ms{]}} &
        \textbf{Gain (CPU)} &
        \textbf{Gain (GPU)} \\
    
    \midrule
    
    \textbf{Cora} &
        244.4 & 
        11.1 & 
        &
        7.2 &
        376.3 &
        &
        0.246 &
        11,022.0 &
        994.8$\times$ &
        29.3$\times$ \\
        
    \textbf{CiteSeer} &
        244.3 & 
        13.6 & 
        &
        10.1 &
        330.0 & 
        &
        0.294 &
        11,318.6 &
        831.2$\times$ &
        34.3$\times$ \\
    
    \textbf{PubMed} &
        362.4 & 
        54.4 & 
        &
        4.8 &
        4,099.5 &
        &
        1.617 &
        12,193.2 &
        224.1$\times$ &
        3.0$\times$ \\
    
    \textbf{Flickr} &
        475.4 & 
        187.8 & 
        &
        14.5 &
        6,146.2 & 
        &
        7.227 &
        12,350.0 &
        65.8$\times$ &
        2.0$\times$ \\
    
    \textbf{Reddit} &
        953.3 & 
        244.4 & 
        &
        171.0 &
        1,362.0 & 
        &
        24.6 &
        9,463.6 &
        38.7$\times$ &
        6.9$\times$ \\
    
    \textbf{Yelp} &
        760.8 & 
        942.2 & 
        &
        110.9 &
        6461.6 & 
        &
        57.5 &
        12,471.7 &
        13.2$\times$ &
        1.9$\times$ \\
    
    \midrule
    
    \textbf{Average} &
    506.8 & 
    242.2 & 
    &
    53.1 &
    3,129.3 &
    &
    15.2 & 
    11,469.9 &
    $\mathbf{361.1\times}$ &
    $\mathbf{12.9\times}$ \\
        
    \bottomrule
    \end{tabular}%
}
\label{tab:latency_results}
\end{table*}

\subsection{Mixed-Precision Arithmetic}
The DegreeQuant algorithm was used to assign the precision for each node in the graph datasets, by stochastically protecting nodes according to their degree (see Section \ref{sec:background}). As shown in Table \ref{tab:datasets}, the ratio of protected nodes is below 3\% for all datasets, suggesting a similar ratio of resources on the accelerator should be allocated to float. Configuration parameters were then chosen as follows; given two node groups, for float and int8 nodes, a resource budget $R^{max, r}_{p}$ (where $p \in [\text{float, int8}]$ is the numerical format and $r \in [\text{LUT, FF, BRAM, DSP}]$ is the resource type) is allocated to each group using the ratio obtained from DegreeQuant. A single-arithmetic variant was synthesized for float and int8, and the resource utilization per nodeslot $C^{r}_{p}$ was estimated for each precision and resource type. Finally, the number of nodeslots $N_p$ for each precision is determined as shown in equation \ref{eq:dq_nodeslots}, where the brackets denote rounding up to the nearest integer.
\begin{equation}\label{eq:dq_nodeslots}
    N_p = \left \lceil \underset{r}{\min} \frac{R^{max, r}_{p}}{ C^{rt}_{p}} \right \rceil
\end{equation}

It was expected that at lower ratios of protected nodes, resources can be distributed across a higher number of nodeslots, due to the lower resource usage of fixed-point cores. In fact, it was found that allocating a single nodeslot to floating-point is normally enough to meet the precision requirement for task accuracy while maximising hardware node parallelism.

\subsection{Performance Analysis}

Each model was first benchmarked on the Intel Xeon CPU and RTX A6000 GPU across all datasets, with randomly initialized node features and layer weights. In each case, the mean latency was obtained over 100 trials to account for runtime jitter due to non-deterministic processes. The GPU cache was emptied prior to each prediction step such that latency readings include off-chip memory access for features and weights. GPU warm-up time was not included, meaning inference times are taken after driver initialization is complete. Finally, inference latency on AGILE was obtained from Modelsim 19.2 simulation results at a frequency of 200MHz, obtained for the Alveo U280 card using the Vivado 23.1 toolflow. As shown in Figure \ref{fig:speedup}, AMPLE led to an improvement in mean inference time compared to the CPU/GPU baselines across all models. Table \ref{tab:latency_results} shows the obtained values for latency and node throughput for GCN.
\section{Conclusion} \label{sec:conclusion}
\vspace{-1mm}
This work presented AMPLE, an FPGA accelerator for GNN inference over large graphs. An event-driven programming flow was introduced, coupled with a dynamic resource allocation mechanism through on-chip network communication, overcoming the performance bottleneck associated with node batching in graphs with non-uniform distribution of node degrees. Using a node-centric data prefetcher, we alleviate the requirement for on-chip storage of weights and activations, enabling GNN acceleration over social media graph datasets. These factors led to an average speedup of $243\times$ and $7.2\times$ compared to CPU and GPU baselines. Finally, we provide the first platform to accelerate graphs quantized at node granularity, demonstrating an optimal resource mapping to maximise node parallelism at a low cost to model accuracy.

\newpage

\bibliography{main}
\bibliographystyle{conference}

\newpage

\appendix

\end{document}